\documentstyle[12pt]{article}
\def \vslash{\!\!\not{\! v}}
\begin{document}
\renewcommand{\thefootnote}{\fnsymbol{footnote}}
\begin{titlepage}
\begin{flushright}
CPT-94/P.3116\\
IJS-TP-95/5\\
1995\\
\end{flushright}
\vspace{.5cm}
\begin{center}
{\Large \bf Saturation of counterterms by resonances in
$K \rightarrow \pi e^{+} e^{-}$ decays
\\}
\vspace{.5cm}
{\large \bf S. Fajfer \\}
\vspace{.5cm}
{\it Centre de Physique Theorique - Section II - CNRS - Luminy, case 907,
F-13288 - MARSEILLE CEDEX 09 - France}\\
and\\
{\it ``J. Stefan'' Institut, University of Ljubljana,
61111 Ljubljana, Slovenia\\}
\vspace{.5cm}
\end{center}
\centerline{\large \bf ABSTRACT}
\vspace{0.5cm}

The decays $K^{+} \rightarrow \pi^{+} e^{+} e^{-}$,
$K_{S} \rightarrow \pi^{0} e^{+} e^{-}$
and $K_{L} \rightarrow \pi^{0} e^{+} e^{-}$ are reinvestigated
within the framework of chiral perturbation theory.
The counterterms induced by strong, electromagnetic and weak
interactions are determined assuming the resonance exchange.
The weak deformation model,
the factorization model and the large $N_{c}$ limit are used to
create a weak Lagrangian.
It is found that the results of the first two approaches depend on the
$H_{1}$ coupling,
defined in the effective chiral Lagrangian of the
${\it O} (p^{4})$ order.
The set of parameters used in the extended Nambu and Jona-Lasinio model can
accommodete
$K^{+} \rightarrow \pi^{+} e^{+} e^{-}$ decay rate
within the factorization approach.
The CP violating  $K_{L} \rightarrow \pi^{0} e^{+} e^{-}$
decay rate is discussed.

\end{titlepage}

\setlength {\baselineskip}{0.75truecm}
\parindent=3pt  
\setcounter{footnote}{1}    

\newcommand{\tr}{\mbox{\rm Tr\space}}
\def \vslash{\!\!\not{\! v}}
\renewcommand{\thefootnote}{\arabic{footnote}}
\setcounter{footnote}{0}
\vspace{.5cm}
{\bf 1 Introduction}\\

The decays $K \rightarrow \pi e^{+} e^{-} $
inspire
many theoretical and experimental studies due to possibility to observe
CP violation
\cite{EPR1,EPR2,EPR3,EPR4,DF}.
There are following possible decays:
$K^{+} \rightarrow \pi^{+} e^{+} e^{-} $,  $K_{S} \rightarrow \pi^{0} e^{+}
e^{-} $ and
$K_{L} \rightarrow \pi^{0} e^{+} e^{-} $.
The CP conserving  $K^{+} \rightarrow \pi^{+} e^{+} e^{-} $
and  $K_{S} \rightarrow \pi^{0} e^{+} e^{-} $  are dominated by virtual
photon exchange. The amplitudes of order
$\it{O}(p^{2})$ vanish at the tree level
\cite{EPR2} and therefore
the leading amplitudes for these transitions are of  $\it{O}(p^{4})$ order
in the chiral perturbation theory (CHPT). At this order there are both one-loop
contributions and tree level contributions.
The one-loop contribution was determined by G. Ecker, A. Pich and E. de Rafael
\cite{EPR1,EPR2,EPR3,EPR4}, but the tree level contribution (or better the
counterterm contribution) motivates  many studies  \cite{DF,AEIN}.
The decay $K_{L} \rightarrow \pi^{0} e^{+} e^{-} $ proceeding via
virtual $\gamma^{*}$, is forbidden in the limit of CP conservation.
The CP conserving process proceeds through two photon exchanges.
A CP violating term is proportional to the known
$\epsilon$ parameter \cite{EPR3}. In addition,
there is a direct $\Delta S = 1$ CP violating effect \cite{DF,BLMM}.
The amplitudes for processes  $K^{+} \rightarrow \pi^{+} \gamma \gamma $
and $K_{S} \rightarrow \pi^{0} \gamma \gamma $  involve the same
counterterm couplings
as processes with one virtual photon exchange.
The effective theory contains a large number of unknown parameters.
The phenomenological parameters appearing in the strong sector can all
be determined, while many of the weak couplings cannot be
fixed by experiment.\\
The authors of ref. \cite{EGPR} have considered
the resonance contribution to the coupling constants of the
${\it O}(p^{4})$ effective chiral Lagrangian.
They find clear evidence for the importance of vector
meson contributions, which account for the bulk of the
low-energy coupling constants.
The same idea
has been questioned in the weak interactions
sector \cite{EKW,IP,SF1,KMW,KMW1}. The authors in
\cite{EKW,KMW,KMW1} have studied
the most general case
of the $SU(3) \times SU(3)$ invariant chiral Lagrangian
using symmetry principles only, and they found that there
appear $37$ independent terms.
These couplings  cannot be determined by present experiments.
Therefore, additional assumptions are needed to describe the weak interactions.
There are two procedures available: \\
a) the ``weak deformation model''\\
b) the ``factorization model''.\\
Both models can be formulated without any reference to resonances.
Because the strong couplings of the ${\it O}(p^{4})$  order in
the chiral Lagrangian
seem to be saturated by resonance exchange \cite{EGPR}, it seems natural
that the weak interactions at ${\it O}(p^{4})$ order in the
chiral Lagrangian might be explained by  resonance contributions.\\
The purpose of this paper is to clarify the role of resonances
in the counterterms for $K \rightarrow \pi e^{+} e^{-} $ decays. \\

The paper is organized as follows: In Sect. 2 we shall
repeat the general results
for $K\rightarrow \pi \gamma^{*}$ and $K\rightarrow \pi \gamma \gamma$
amplitudes and we explain the electroweak counterterm-couplings
by resonance exchange.
In Sect. 3 we shall analyse these terms using "factorization model",
"weak deformatoion models" and we shall apply the extended Nambu and
Jona-Lasinio
model
for the couplings in the strong chiral Lagrangian of ${\it O}(p^{4})$ order.
In Sect. 4 the
large $N_{c}$ model is combined with the resonance
saturation.  The $K_{L} \rightarrow \pi^{0} e^{+} e^{-}$ decay rate
is briefly analysed in Sect. 5.
In Sect. 6 we give summary of our results.\\

{\bf 2 $K\rightarrow \pi \gamma^{*}$
and $K\rightarrow \pi \gamma \gamma$ decays in CHPT}\\

It was shown that in the chiral perturbation theory at
${\it O}(p^{2})$ $K\rightarrow \pi \gamma^{*}$ transitions
are forbidden for a virtual photon
$\gamma^{*}(q)$ for any value of $q^{2}$ \cite{EPR1}.
Combining the contributions coming from one-loop and the counterterms,
induced by strong, weak and electromagnetic interactions,
\cite{EPR2,EPR3}, the amplitudes for $K^{+} \rightarrow \pi^{+} \gamma^{*}$
and  $K_{S} \rightarrow \pi^{0} \gamma^{*}$  can be written as
\begin{eqnarray}
A(K^{+} \rightarrow \pi^{+} \gamma^{*}) & = & \frac{G_{8} e}{(4 \pi)^{2}}
q^{2} \Phi_{+}(q^{2}) \epsilon^{\mu} (p^{\prime} + p)_{\mu}
\label{e1}
\end{eqnarray}
\begin{eqnarray}
A(K^{0}_{S} \rightarrow \pi^{0} \gamma^{*}) & = & \frac{G_{8} e}{(4 \pi)^{2}}
q^{2} \Phi_{S}(q^{2}) \epsilon^{\mu} (p^{\prime} + p)_{\mu}
\label{e2}
\end{eqnarray}
where  $p$ and $p^\prime$ are pion's  and  kaon's momenta,
$G_{8} = {\sqrt \frac{1}{2}}  G_{F} s_{1} c_{1} c_{3} g_{8}$ is defined in
\cite{EPR1} and from $K\rightarrow \pi \pi$ it was found that $|g_{8}| = 5.1$.
The calculated one-loop correction reduces this value to $|g_{8}^{loop}| = 4.3$
\cite{DF,KMW1}. The factors $\Phi_{+,S}$ are
split into one-loop and counterterm contributions
\begin{eqnarray}
\Phi_{+} & = & W_{+} + \Phi_{K} + \Phi_{\pi}
\label{e3}
\end{eqnarray}
\begin{eqnarray}
\Phi_{S} & = & W_{S} + 2 \Phi_{K}
\label{e4}
\end{eqnarray}
The loop contributions $\Phi_{K}$ and $\Phi_{\pi}$ are determined in
\cite{EPR1,EPR2,EPR3}.
The $W_{+,S}$ have been defined as \cite{EPR2,EPR3,EPR4}
\begin{eqnarray}
W_{+} & = & -\frac{16}{3} \pi^{2} [ W_{1}^{r} + 2 W_{2}^{r} - 12
L_{9}^{r}(\mu)]
+ \frac{1}{3} log\frac{\mu^{2}}{M_{K} M_{\pi}}
\label{e5}
\end{eqnarray}
and
\begin{eqnarray}
W_{S} & = & -\frac{16}{3} \pi^{2} [ W_{1}^{r} - W_{2}^{r}]
+ \frac{1}{3} log\frac{\mu^{2}}{M_{K}^{2}}
\label{e6}
\end{eqnarray}
In these equations $L_{9}^{r}(\mu)$ is a coupling constant at
${\it O}(p^{4})$ strong Lagrangian defined in \cite{EGPR,GL,BC}
\begin{eqnarray}
{\cal L}_{4} & = & ...-i Tr L_{9}( f^{\alpha \beta}_{+} u_{\alpha} u_{\beta})
 +  \frac{1}{4} (L_{10} + 2 H_{1}) Tr (f_{+\alpha \beta}f^{\alpha
\beta}_{+})\nonumber\\
& - & \frac{1}{4} (L_{10} -  2 H_{1}) Tr (f_{-\alpha \beta}f^{\alpha
\beta}_{-})
\label{i6}
\end{eqnarray}
with $u^{2} = U = exp (\frac{i}{f}\Sigma_{i} \lambda^{i} \phi_{i}) $ and
$f\simeq f_{\pi} = 0.0933 GeV$.
The notation here is defined below:
\begin{eqnarray}
f^{\alpha \beta}_{\pm} & = & u F^{\alpha \beta}_{L} u ^{\dag} \pm u^{\dag}
F^{\alpha \beta}_{R}
u\label{i6a}
\end{eqnarray}
\begin{eqnarray}
F^{\alpha \beta}_{L} & = & \partial^{\alpha} l^{\beta} - \partial^{\beta}
l^{\alpha}
- i [l^{\alpha}, l^{\beta}]\label{e6b}
\end{eqnarray}
\begin{eqnarray}
F^{\alpha \beta}_{R} & = & \partial^{\alpha} r^{\beta} - \partial^{\beta}
r^{\alpha}
- i [r^{\alpha}, r^{\beta}]\label{e6c}
\end{eqnarray}
\begin{eqnarray}
u_{\alpha} & = & i u^{\dag} D_{\alpha} U u^{\dag}\label{e6d}
\end{eqnarray}
\begin{eqnarray}
D_{\alpha} & = & \partial _{\alpha} U - i r_{\alpha} U + i U
l_{\alpha}\label{e6e}
\end{eqnarray}
and $l_{\alpha}$ and $r_{\alpha}$ denote left and right matrix external fields,
with spin $1$. In the presence of the external electromagnetic fields only
$l_{\alpha} = r_{\alpha} = |e| Q A_{\alpha}$.\\
The coupling constant $L_{9}$ is connected with the renormalized $L_{9}^{r}$ as
\begin{eqnarray}
L_{9} & = & L_{9}^{r} + \Gamma _{9} \frac{\nu^{-\epsilon}}{(4 \pi)^{2}
{\hat{\epsilon}}},
\enspace \Gamma_{9} = \frac{1}{4}. \label{f6}
\end{eqnarray}
In this paper we choose $\nu = m_{\rho}$ for the renormalization scale $\nu$.
In a similar way
$W_{1}$ and $W_{2}$ counterterms of weak and electromagnetic origin can be
written in the form \cite{EPR2}
\begin{eqnarray}
W_{1} & = & W_{1}^{r} + \Gamma _{9} \frac{\nu^{-\epsilon}}{(4 \pi)^{2}
{\hat{\epsilon}}},\enspace
\Gamma_{9} = \frac{1}{4}\label{f6a}
\label{f7}
\end{eqnarray}
In ref.  \cite{EPR2,EPR3} it was found
that in the case of   $K\rightarrow \pi \gamma^{*}$  and $K\rightarrow \pi
\gamma \gamma$
decays
there are three relevant local counterterms:
\begin{eqnarray}
{\cal L}_{4}^{\Delta S = 1,em} & = & \frac{i e G_{8} f^{2}}{2} F^{\mu \nu}
[ W_{1} Tr (Q \lambda_{6-i7} U^{\dag} D_{\mu}U U^{\dag} D_{\nu}U)\nonumber\\
 & + & W_{4} Tr (Q  U^{\dag} D_{\mu}U \lambda_{6-i7}
U^{\dag} D_{\nu} U)]\nonumber\\
& + &\frac{e ^{2} f^{2}}{2} G_{8} F^{\mu \nu} F_{\mu \nu}
 W_{4} Tr (\lambda_{6-i7} Q U^{\dag}Q U) + h.c.\label{e7}
\end{eqnarray}
The analysis of $K^{+}\rightarrow \pi^{+}\gamma \gamma$ decay
shows that there is the following combination of coupling constants
\begin{eqnarray}
\hat{c} & = & 32 \pi^{2} [ 4 (L_{9} + L_{10}) - \frac{1}{3} (W_{1} +
2 W_{2} + 2 W_{4})].\label{e8}
\end{eqnarray}
The combination $L_{9} + L_{10}$
and the weak contribution  $W_{1} + 2 W_{2} + 2 W_{4}$ are
a renormalization scale invariant \cite{EPR3,GL}.
The loop contribution is finite in $K\rightarrow \pi\gamma \gamma$ decays
and for $K^{0}\rightarrow \pi^{0}\gamma \gamma$ decay
${\it O}(p^{4})$ couterterms
contribution vanishes.\\
In order to determine the  resonance saturation of
the counterterm couplings  in the
chiral Lagrangian  at ${\it O}(p^{4})$,
the lowest order couplings in the chiral expanssion are needed.
All these couplings are of ${\it O}(p^{2})$, and resonance exchange will
automatically produce contributions of ${\it O}(p^{4})$.
Using the equations of motion for resonances, the authors of ref. \cite{EGPR}
have found that $L_{9}$, $L_{10}$ and $H_{1}$
get contributions from vector and axial-vector reconances. They are
\begin{eqnarray}
L_{9}^{V} & = & \frac{F_{V}G_{V}}{2 M_{V}^{2}}, \enspace
L_{10}^{V} = -  \frac{F_{V}^{2}}{4 M_{V}^{2}},  \enspace  \nonumber\\
H_{1}^{V}&  = &- \frac{F_{V}^{2}}{8 M_{V}^{2}}
\label{e9}
\end{eqnarray}
\begin{eqnarray}
L_{9}^{A} & = & 0, \enspace
L_{10}^{A} =  \frac{F_{A}^{2}}{4 M_{A}^{2}}, \enspace
H_{1}^{V}  = - \frac{F_{A}^{2}}{8 M_{A}^{2}}
\label{e10}
\end{eqnarray}
where $M_{V}$ and $M_{A}$ are the octet masses of
vector and axial-vector mesons.
The octet couplings $F_{V}$ and $G_{V}$ can in principle be determined from
the decay rates $\rho \rightarrow l^{+}l^{-}$ and $\rho \rightarrow 2 \pi$,
respectively, and they are
\begin{equation}
|F_{V}| = 0.154\enspace GeV, \enspace |G_{V}| = 0.069\enspace GeV. \label{e11}
\end{equation}
For the determination of $F_{A}$ and the octet mass $M_{A}$ in the chiral
limit the authors of \cite{EGPR} have
used Weinberg's sum rules \cite{W}.
These two sum rules connect
axial vector meson with vector meson parameters,
\begin{equation}
F_{V}^{2} = F_{A}^{2} + f^{2}
\label{e12}
\end{equation}
\begin{equation}
M_{V} ^{2}F_{V}^{2} = M_{A}^{2} F_{A}^{2},
\label{e13}
\end{equation}
 \begin{equation}
F_{A}^{2} = 0.128\enspace GeV
\label{e14}
\end{equation}
\begin{equation}
M_{A}^{2} = 0.968\enspace GeV
\label{e15}
\end{equation}
The relevant ${\it O}(p^{4})$ couplings
$L_{9}$, $L_{10}$ and $H_{1}$  are in general divergent, like the rest of $
{\it O}(p^{4})$ couplings. They depend on the renormalization scale $\nu$
which is not seen in
the observable quantities. The results we use, like \cite{EGPR,GL,BC}
are obtained at $\nu = M_{\rho}$.
\begin{equation}
L_{9}^{r} = 6.9 \times 10^{-3} \enspace
\label{e16a}
\end{equation}
\begin{equation}
L_{10}^{r} = -6.0 \times 10^{-3} \enspace
\label{e16b}
\end{equation}
\begin{equation}
H_{1}^{r} = 7.0\times 10^{-3}
\label{e16c}
\end{equation}
The $H_{1}^{r} $ is not accessible experimentally, but vector and
axial vector mesons determine this coupling at scale $\nu = M_{\rho}$
and, once the form of strong chiral Lagrangian at ${\it O}(p^{4})$
is chosen in the form \cite{EGPR,GL,BC}, this term is fixed by resonance
exchange.\\

{\bf 3 Counterterms and models for the
weak Lagrangian at ${\it O}(p^{4})$ order}\\

The weak deformation model and
the factorization model are used in the literature \cite{EPR3,EKW}
towards constructing the weak Lagrangian
of ${\it O}(p^{4})$ order.
Both models rest on the assumption that the strong chiral
Lagrangian already determines the dominant features of
$\Delta S = 1$ effective Lagrangian. It is obvious that such a procedure
cannot  be expected to yield the complete weak Lagrangian,
since the short-distance
contribution has no equivalence in the strong sector.
One might expect that the dominant long-distance contribution results
from resonances exchange like in the strong sector.
We remark that there are two possible ways to obtain
the resonance contribution to ${\cal L}_{w}$:
One might first calculate the resonance contributions to the strong Lagrangian
and then use a weak deformation or factorization procedure.
Or, one might apply the weak deformation or factorization to
the strong resonance Lagrangian of lowest order first and then
integrate out the resonances,
using their equations of motion. If the second approach is used, the
resonance contribution recognized in  $H_{1}$ counterterm of the Lagrangian
(\ref{i6}), is undoubtfully present. \\

a) Weak deformation model \\

This model is inspired by the geometry of the
coset space $G/SU(3)_{V}$ \cite{EPR4}.
It can easily be obtained starting with the
strong Lagrangian of ${\it O}(p^{2})$
\begin{eqnarray}
{\cal L}_{2} & = & \frac{f^{2}}{4} Tr (u_{\mu} u^{\mu}),\label{e17}
\end{eqnarray}
and making the replacement
\begin{eqnarray}
u_{\mu} & \rightarrow & u_{\mu} + G_{8} f^{2}\{ u_{\mu}, \Delta \}
- \frac{2}{3} G_{8} f^{2} Tr (u_{\mu} \Delta), \label{e18}
\end{eqnarray}
where
\begin{eqnarray}
\Delta  & = & u \lambda_{6} u^{\dag}.\label{e19}
\end{eqnarray}
It is useful to introduce two forms
\begin{eqnarray}
l_{\mu}  & = & u [ \partial - i (v_{\mu} - a_{\mu}) ]u^{\dag}
= \Gamma _{\mu} + \frac{1}{2} i u_{\mu}\label{e20a}
\end{eqnarray}
\begin{eqnarray}
r_{\mu}  & = & u [ \partial - i (v_{\mu} + a_{\mu}) ]u^{\dag}
= \Gamma _{\mu} - \frac{1}{2} i u_{\mu}.\label{e20b}
\end{eqnarray}
In this model $\Gamma_{\mu}$ is deformed to
\begin{eqnarray}
\Gamma_{\mu} & \rightarrow & \Gamma_{\mu} +\frac{1}{2}i G_{8} f^{2}_{\pi}
\{ u_{\mu}, \Delta \}
- \frac{1}{3}i G_{8} f^{2}_{\pi} Tr (u_{\mu} \Delta) \label{e21}
\end{eqnarray}
After performing the weak deformation on ${\cal L}_{4}$ in (\ref{i6})
the counterterm coupling defined in (\ref{e7})
are found to be
\begin{eqnarray}
W_{1}^{r} & = & 4 (L_{9}^{r} + L_{10}^{r} + 2 H_{1}^{r})
\label{e22a}
\end{eqnarray}
\begin{eqnarray}
W_{2}^{r} & = & 4 L_{9}^{r}
\label{e22b}
\end{eqnarray}
\begin{eqnarray}
W_{4}^{r} & = & 4 (L_{10}^{r} - H_{1}^{r}).
\label{e22c}
\end{eqnarray}
With the numerical results for $L_{9}^{r}$, $L_{10}^{r}$ and
$H_{1}^{r}$ from (\ref{e16a}),
(\ref{e16b}) and (\ref{e16c}) we find
\begin{eqnarray}
W_{1}^{WDM} & = & - 0.0524
\label{e29a}
\end{eqnarray}
\begin{eqnarray}
W_{2}^{WDM} & = & 0.0276
\label{e29b}
\end{eqnarray}
\begin{eqnarray}
W_{4}^{WDM} & = & 0.004
\label{e29c}
\end{eqnarray}
what leads, using (\ref{e5}) and (\ref{e6}), to $W_{+}^{WDM}  = -5.01$
$W_{S}^{WDM} = 4.50$.
Among three possible decays only the decay rate of
$K^{+} \rightarrow \pi^{+} e^{+} e^{-}$
has been observed. The experimental bound for the branching ratio
$K^{+} \rightarrow \pi^{+} e^{+} e^{-}$ is obtained
from Brookhaven experiment \cite{BE}
\begin{equation}
BR(K^{+} \rightarrow \pi^{+} e^{+} e^{-}) = (2.99\pm 0.22)\times 10^{-7},
\label{e31}
\end{equation}
resulting in  bounds $W_{+}^{WDM}  = 0.89_{-0.14}^{+0.24} $.
It is obvious that the numerical value obtained by using
the weak deformation model does not explain this experimental limit.\\
In the case of $K^{+} \rightarrow \pi^{+} \gamma \gamma$
the counterterm coupling defined in (\ref{e8}) is
${\hat c}^{WDM} = 0$.
In ref. \cite{DF} the relation (\ref{e22b}) was criticized as a result of
assumption which is not part of the CHPT. We point out
that this result originally was derived using the octet dominance hypothesis.
Namely, the authors of ref. \cite{EPR1} have noticed that the one-loop
amplitudes
for $K^{+}\rightarrow \pi^{+} \gamma^{*}$,
$K^{0}\rightarrow \pi^{0} \gamma^{*}$ and
$\eta\rightarrow \bar{K}^{0} \gamma^{*}$, if
$m_{K} = m_{\pi}$, satisfy the relation
\begin{eqnarray}
A(K^{+}\rightarrow \pi^{+} \gamma^{*})|_{l}& = &
- \sqrt{2}  A(K^{0}\rightarrow \pi^{0} \gamma^{*})|_{l} =
\sqrt{\frac{2}{3} }A(\eta\rightarrow \bar{K}^{0} \gamma^{*})|_{l}.
\label{e22d}
\end{eqnarray}
This relation restricts two pseudoscalars to be in the pure octet in the
$SU(3)$ limit. We find that without $SU(3)$ symmetry these amplitudes satisfy
\begin{eqnarray}
A(K^{+}\rightarrow \pi^{+} \gamma^{*})|_{l}& = &
-\frac{1}{\sqrt{2}} ( A(K^{0}\rightarrow \pi^{0} \gamma^{*})|_{l} +
\sqrt{\frac{1}{3} }A(\eta\rightarrow \bar{K}^{0} \gamma^{*})|_{l}).
\label{e22e}
\end{eqnarray}
Generally, as it was noticed in ref. \cite{EPR1},
it means that two pseudoscalars can be in state
of representation $8$, $10$, and $\bar{10}$ of $SU(3)$, and with the
help of Clebsch-Gordon coefficients for $8\times 10$ of $SU(3)$ group
it can be seen that decuplet components  cancel out in (\ref{e22e}).
Imposing (\ref{e22d}), if $SU(3)$ limit holds,  or (\ref{e22e})
when $SU(3)$ is broken, on the
amplitudes containing the weak counterterms, we find that
relation (\ref{e22b}) exists in both cases, with and without
$SU(3)$ limit.
Athough our result for $W_{+}$ disagrees with the experimental bound,
we cannot rule out the weak deformation model. The model itself contains
very important feature: the relations
(\ref{e22a}), (\ref{e22b})  and (\ref{e22c}) are scale independent
\cite{EPR4} and further study of this model would be useful.
\vspace{1cm}

b) Factorization model\\

The effective $\Delta S = 1$ weak Hamiltonian is given by
\begin{eqnarray}
{\it H}_{eff} & = & \frac{G}{{\sqrt 2}} C_{i}(\mu^{2}) Q_{i} + h.c.
\label{e23}
\end{eqnarray}
where $Q_{i}$ are four-quark operators and $C_{i}(\mu^{2})$  are Wilson
coefficients depending on the QCD renormalization scale $\mu$ .
To lowest order in CHPT the left-current is realized as
\begin{eqnarray}
{\it J}_{\mu}^{(1)} & = & \frac{\delta S_{2}}{\delta l_{\mu}},
\label{e24}
\end{eqnarray}
where $S_{2}$ is the action determined by ${\cal L}_{2}$, what gives
\begin{eqnarray}
{\it J}_{\mu}^{(1)} & = & -\frac{f^{2}}{2} u^{\dag}u_{\mu} u \nonumber\\
& = & - i \frac{f^{2}}{2}\{U^{\dag} \partial_{\mu} U + i e A_{\mu} U^{\dag}
[U,Q]\}.
\label{e24a}
\end{eqnarray}
To ${\it O}(p^{4})$, the factorization model is
defined for the dominant octet part
of the weak Lagrangian by
\begin{eqnarray}
{\cal L}_{W4}^{FM} & = & 4 g_{8} Tr (\lambda_{6-i7} \{{\it J}_{\mu}^{(1)},
{\it J}^{{\mu}(3)}\}),
\label{e25}
\end{eqnarray}
where ${\it J}_{\mu}^{(3)}$ is
determined from
\begin{eqnarray}
{\cal J}_{\mu}^{(3)} & = & \frac{\delta S_{4}}{\delta l_{\mu}}.
\label{e26}
\end{eqnarray}
In our case it leads to
\begin{eqnarray}
{\cal J}_{\mu}^{(3)} & = & -L_{9} \{ eF_{\mu \nu} ([L^{\nu}, U^{\dag} Q U]
+ [L^{\nu}, Q])\}\nonumber\\
& + & i L_{9} \partial^{\nu} \{ [L_{\mu},L_{\nu}] + i e A_{\mu}
(-[L_{\nu},Q] + [L_{\nu},U^{\dag} Q U])\}\nonumber\\
& + & 2 L_{10} \partial^{\nu} ( e F_{\mu \nu} U^{\dag} Q U )
+ 4 H_{1} \partial^{\nu} (e F_{\mu \nu} Q)
\label{e27}
\end{eqnarray}
where $L_{\mu} = U^{\dag} \partial_{\mu} U$.
Using these two currents, we easily find
\begin{eqnarray}
W_{1}^{r} & = & 8 (L_{9}^{r} + L_{10}^{r} + 2 H_{1}^{r})
\label{e28a}
\end{eqnarray}
\begin{eqnarray}
W_{2}^{r} & = & 8 L_{9}^{r}
\label{e28b}
\end{eqnarray}
\begin{eqnarray}
W_{4}^{r} & = & 8 (L_{10}^{r} - H_{1}^{r})
\label{e28c}
\end{eqnarray}
With the help of the numerical results for $L_{9}^{r}$, $L_{10}^{r}$ and
$H_{1}^{r}$ from (\ref{e16a}),
(\ref{e16b}) and (\ref{e16c}) we derive
\begin{eqnarray}
W_{1}^{FM} & = & - 0.1048
\label{e30a}
\end{eqnarray}
\begin{eqnarray}
W_{2}^{FM} & = & 0.0552
\label{e30b}
\end{eqnarray}
\begin{eqnarray}
W_{4}^{FM} & = & 0.008,
\label{e30c}
\end{eqnarray}
what results in $W_{+}^{FM}  = -4.87$  and $W_{S}^{FM} = 8.67$. The result
for $W_{+}$, like the one calculated in the case of the weak deformation model,
disagees with the experimental one. The combination of the
counterterm couplings derived in (\ref{e8}) gives
${\hat c}^{FM} = -1.14$.
At present there is only an upper bound for the branching ratio
of this process
$BR (K^{+} \rightarrow \pi^{+} \gamma \gamma) < 1\cdot 10^{-6}$ \cite{A} .
This upper bound and the analysis of ref. \cite{EPR3}
lead to the limits  $-7.02 < {\hat c}< 1.89$ and the obtained value for
${\hat c}^{FM} = -1.14$ is allowed by this bounds.\\

Now, we search for the model which might accommodate
this experimental result and
we apply  the extended Nambu and Jona-Lasinio model
described in ref. \cite{BBR}.
In this reference, the low-energy effective action of an extended
Nambu and Jona-Lasinio model to ${\it O}(p^{4})$ in the chiral counting is
derived.
The couplings of our interest are $L_{9}$, $L_{10}$ and $H_{1}$.
It was found \cite{BBR} that these couplings can be expressed as
\begin{eqnarray}
L_{9} & = & \frac{N_{c}}{16 \pi^{2}} \frac{1}{6}
[ (1 - g_{A}^{2}) \Gamma_{0}(1 + \gamma_{03})  +
2 g_{A}^{2} \Gamma_{1}(1 + \frac{3}{2}\gamma_{12}- \frac{1}{2} \gamma_{13})]
\label{e31a}
\end{eqnarray}
\begin{eqnarray}
L_{10} & = &- \frac{N_{c}}{16 \pi^{2}} \frac{1}{6}
[ (1 - g_{A}^{2}) \Gamma_{0}(1 + \gamma_{03})  +
 g_{A}^{2} \Gamma_{1}(1 + \gamma_{13})]
\label{e31b}
\end{eqnarray}
\begin{eqnarray}
H_{1} & = &- \frac{N_{c}}{16 \pi^{2}} \frac{1}{12}
[ (1 - g_{A}^{2}) \Gamma_{0}(1 + \gamma_{03})  -
 g_{A}^{2} \Gamma_{1}(1 + \gamma_{13})].
\label{e31c}
\end{eqnarray}
In these equations we use
\begin{eqnarray}
\gamma_{03}  & = & -\frac{\Gamma (2,x)}{\Gamma (0,x)} \frac{3}{5} g
\label{e32a}
\end{eqnarray}
\begin{eqnarray}
\gamma_{12}  & = & \frac{\Gamma (3,x)}{\Gamma (1,x)} \frac{1}{5} g
\label{e32b}
\end{eqnarray}
\begin{eqnarray}
\gamma_{13}  & = & -\frac{\Gamma (3,x)}{\Gamma (1,x)} \frac{3}{5} g,
\label{e32c}
\end{eqnarray}
where $\Gamma (n,x)$ denotes the incomplete gamma function
\begin{eqnarray}
\Gamma (n-2, x = \frac{M_{Q}^{2}}{\lambda_{\chi}}) & = &
\int_{\frac{M_{Q}^{2}}{\lambda_{\chi}^{2}}}^{\infty} \frac{d z}{z} e ^{-z}
z^{n-2},
 \enspace n = 1,2,3,...,
\label{e33}
\end{eqnarray}
$g_{A}$ can be identified as a constant of the constituent chiral quark model
and $g$ might be connected to the
gluon vacuum condensate \cite{BBR},
$\lambda_{\chi}$ is a cut-off scale,
$M_{Q}$ is the constituent chiral quark-mass.
For the most favorable set of parameters $g_{A} = 0.61$,
$M_{Q} = 0.265 GeV$ and $\lambda_{\chi} = 1.165 GeV$ \cite{BBR},
it is easy to calculate $L_{9}^{r} = 7.0 \cdot 10^{-3}$,
$L_{10}^{r} = -5.9 \cdot 10^{-3}$ and $H_{1}^{r} = -4.7 \cdot 10^{-3}$.
These values lead to
\begin{eqnarray}
W_{1}^{WDM} & = & -0.0327
\label{e34a}
\end{eqnarray}
\begin{eqnarray}
W_{2}^{WDM} & = & 0.028
\label{e34b}
\end{eqnarray}
\begin{eqnarray}
W_{4}^{WDM} & = & -0.0049
\label{e34c}
\end{eqnarray}
and consequently  $W_{+}^{WDM}  = 3.91$, $W_{S}^{WDM} = 6.69$ and
${\hat c}^{WDM}  = 0$.
For the factorization model the  values of $W_{i}$ are doubled
and physically relevant values are $W_{+}^{FM}  = 2.69$,
$W_{S}^{FM} = 6.69$ and ${\hat c}^{FM} = -1.41$.
The numerical values of $W_{i}$ are calculated at the scale
$\nu = m_{\rho}$.
In Table 1 we present $W_{+}$, $W_{S}$
and  ${\hat c}$ counterterms calculated for some of the values of
the scale $\nu$.
Analysing fits from ref. \cite{BBR} we find that their  fit. $4$ gives
most favorable value of the $W_{+}$. In this fit there is no explicit vector
(axial-vector)  degree of freedom.
Using this fit, it was found  $L_{9}^{r} = 5.8 \cdot 10^{-3}$,
$L_{10}^{r} = -5.1 \cdot 10^{-3}$ and $H_{1}^{r} = -2.4 \cdot 10^{-3}$.
These values leads to
\begin{eqnarray}
W_{1}^{FM} & = & -0.0328
\label{e36a}
\end{eqnarray}
\begin{eqnarray}
W_{2}^{FM} & = & 0.0464
\label{e36b}
\end{eqnarray}
\begin{eqnarray}
W_{4}^{FM} & = & -0.0216,
\label{e36c}
\end{eqnarray}
resulting in $W_{+}^{FM}  = 1.22$, $W_{S}^{FM} =  4.46$ and ${\hat c}^{FM}
= -0.88$.
for the scale $\nu = m_{\rho}$.
For the scale $\nu = 0.265$ GeV these values are $W_{+}^{FM}  = 0.51$,
$W_{S}^{FM}  = 3.75$ and ${\hat c}^{FM} = -0.88$.
The strenght of the counterterm coupling ${\hat c}^{FM}= -0.88$ is
within the experimental limits.
We can conclude that the factorization model combined with the parameters
used in the extended Nambu and Jona-Lasinio model can accommodate
the experimentally measured
decay rate $BR(K^{+} \rightarrow \pi^{+} e^{+} e^{-})$
and  the bounds on conterterm coupling ${\hat c}$ obtained from
$BR(K^{+} \rightarrow \pi^{+} \gamma \gamma) $ decay rate.\\

{\bf 4 Large $N_{c}$ limit and  counterterms}\\

Using the large $N_{c}$ limit, $\Delta S = 1$ effective Hamiltonian is
given by \cite{BP,PR}
\begin{eqnarray}
{\it H}_{eff}^{\Delta S = 1} & = & \frac{G_{F}}{{\sqrt 2}}
V_{ud} V_{us}^{*} Q_{2}
+ h.c.
\label{e40}
\end{eqnarray}
where $Q_{2} = 4 ({\bar s}_{L} \gamma_{\mu} u_{L})
({\bar u}_{L} \gamma_{\mu} d_{L})$.
The full effective Hamiltonian at the leading order of large $N_{c}$ is
\begin{eqnarray}
{\it H}_{eff}^{\Delta S = 1} & = &
4(G_{8}^{(\frac{1}{2})}{\it H}_{8}^{(\frac{1}{2})} +
G_{27}^{(\frac{1}{2})}{\it H}_{27}^{(\frac{1}{2})} +
G_{27}^{(\frac{3}{2})}{\it H}_{27}^{(\frac{3}{2})}),
\label{e43}
\end{eqnarray}
where in the strict large-$N_{c}$
$g_{8}
^{(\frac{1}{2})}|_{N_{c}\rightarrow \infty} = \frac{3}{5}$,
$g_{27}^{(\frac{1}{2})}|_{N_{c}\rightarrow \infty} = \frac{1}{15}$,
$g_{27}^{(\frac{3}{2})}|_{N_{c}\rightarrow \infty} = \frac{1}{3}$.
In this expression
\begin{eqnarray}
{\it H}_{27} & = & -\frac{2}{3} ({\it L}_{\mu})_{21}({\it L}^{\mu})_{13}
- ({\it L}_{\mu})_{23}({\it L}^{\mu})_{11}.
\label{e44}
\end{eqnarray}
${\it H}_{27}$ induces both $|\Delta I| = \frac{1}{2}$ and
$|\Delta I| = \frac{3}{2}$ transitions via its components
\begin{eqnarray}
\it{H}_{27} & = & \frac{1}{9} \it{H}_{27}^{(\frac{1}{2})}
+ \frac{5}{9} \it{H}_{27}^{(\frac{3}{2})}.
\label{e45}
\end{eqnarray}
In our calculation we kept both components of
$Q_{2}$ with $|\Delta I| = \frac{1}{2}$ and
$|\Delta I| = \frac{3}{2}$,
the octet and twenty-sevenplet .
With the use of the expressions (\ref{e24}) and (\ref{e27}) we derive
\begin{eqnarray}
A(K^{+} \rightarrow \pi^{+} \gamma^{*} ) & = &
{\sqrt \frac{1}{2}}  G_{F} s_{1} c_{1} c_{3} 4 L_{9} q^{2}
\epsilon^{\mu} (p^{\prime} + p)_{\mu}
\label{e40a}
\end{eqnarray}
\begin{eqnarray}
A(K^{0} \rightarrow \pi^{0} \gamma^{*} ) & = & 0.
\label{e40b}
\end{eqnarray}
For the  $K^{+} \rightarrow \pi^{+} \gamma \gamma$ decay we obtain
\begin{eqnarray}
G_{8} {\hat c}& = & - 32 \pi^{2}{\sqrt \frac{1}{2}}  G_{F} s_{1} c_{1} c_{3}
4 (L_{9} + L_{10})\label{e40x}.
\end{eqnarray}
The authors of ref. \cite{BP} have calculated
the octet component of the isospin
$\frac{1}{2}$ effective Hamiltonian. In the large $N_{c}$ limit the $G_{8}$
coupling is $g_{8} = \frac{3}{5}$
(to be compared with the experimental value 5.1)
They found that the factorization of the $G_{8}W_{i}$,  where $i = 1,2,4$,
calculated at order ${\it O}(p^{4})$ of the chiral Lagrangian is not
valid at order $N_{c}$ in the $\frac{1}{N_{c}}$- expansion. We confirm
their result
\begin{eqnarray}
W_{1} & = & W_{2}
\label{e41a}
\end{eqnarray}
\begin{eqnarray}
G_{8} W_{2} & = & 8 G_{8}|_{\frac{1}{N_{c}}} L_{9}
\label{e42b}
\end{eqnarray}
\begin{eqnarray}
G_{8} W_{4} & = & 12 G_{8}|_{\frac{1}{N_{c}}} L_{10}.
\label{e42c}
\end{eqnarray}
where the subscript $\frac{1}{N_{c}}$ explains that the given coupling constant
is
determined at the leading order of the $\frac{1}{N_{c}}$ expansion.
We derive numerical results, taking into account the values of
$L_{9}$ and $L_{10}$ when they are dominated by resonance exchange (\ref{e16a})
and (\ref{e16b}).
They correspond to the expressions of $G_{8}W_{i}$
\begin{eqnarray}
G_{8}W_{1} & \rightarrow & {\sqrt \frac{1}{2}}  G_{F} s_{1} c_{1} c_{3}0.0552
\label{e46a}
\end{eqnarray}
\begin{eqnarray}
G_{8}W_{2} &\rightarrow  & {\sqrt \frac{1}{2}}  G_{F} s_{1} c_{1} c_{3}0.0552
\label{e46b}
\end{eqnarray}
\begin{eqnarray}
G_{8}W_{4} & \rightarrow & -{\sqrt \frac{1}{2}}  G_{F} s_{1} c_{1} c_{3} 0.072
\label{e46c}
\end{eqnarray}
or $G_{8}W_{S}\rightarrow  - {\sqrt \frac{1}{2}}
G_{F} s_{1} c_{1} c_{3}0.237$,
and $G_{8}W_{+} \rightarrow - {\sqrt \frac{1}{2}}
G_{F} s_{1} c_{1} c_{3}0.737$.
The result corresponding  to $G_{8}W_{+}$ is
quite close to the lower experimental bound.
The result which corresponds to
${\sqrt \frac{1}{2}}  G_{F} s_{1} c_{1} c_{3}{\hat c} $ is
also within the experimental bound $G_{8} {\hat c} =
- {\sqrt \frac{1}{2}}  G_{F} s_{1} c_{1} c_{3} 1.01$ .
The analysis of ref. \cite{BP} contains the effective the Hamiltonian
calculated at the order $p^{4}$ in the chiral
expansion and at the order $N_{c}$
in the $\frac{1}{N_{c}}$-expansion.
Their result leads to the conclusion that
the factorization in $G_{8} W_{i}$ is not
valid when next-to-leading corrections in $\frac{1}{N_{c}}$ are calculated.
Here, we have modified the leading term in $\frac{1}{N_{c}}$-expansion
taking into account the contribution of $27$ component of the $Q_{2}$ operator.
Results are in slightly better agreement (factor of $\frac{5}{3}$ increase)
than the ones derived in \cite{BP} for their leading
term in $\frac{1}{N_{c}}$-expansion.
It is interesting in this approach the counterterm contributions do not
depend on $H_{1}$ when calculated at the leading order in $\frac{1}{N_{c}}$.\\

{\bf 5 $K_{L} \rightarrow \pi^{0} e^{+} e^{-}$ and CP violation}\\

The decay  $K_{L} \rightarrow \pi^{0} e^{+} e^{-}$ is being investigated
as a signal of direct $\Delta S = 1$ CP violation. In addition to a
CP conserving
process, which proceeds through two photon exchanges, there are two
kinds of the CP violating decay: one proportional
to the well known parameter $\epsilon$  and the other
direct CP violating effect. The direct CP violating component is the simplest.
Uncertainties are coming from
poor knowledge of the Standard model parameters,
top quark mass and CKM matrix elements.
The QCD short distance corrections to
this mode have been analyzed to next-to-leading
order by Buras et al. \cite{BLMM}.
The weak operator responsible for this decay is
\begin{eqnarray}
{\cal H}_{w} & = & \frac{G_{F}}{{\sqrt 2}}[C_{7V} + C_{7A}],
\label{e47}
\end{eqnarray}
where
\begin{eqnarray}
Q_{7V} & = & ({\bar s} d)_{V-A} ({\bar e} e)_{V}
\label{e48}
\end{eqnarray}
\begin{eqnarray}
Q_{7A} & = & ({\bar s} d)_{V-A} ({\bar e} e)_{A}.
\label{e49}
\end{eqnarray}
The amplitude for $K_{L} \rightarrow \pi^{0} e^{+} e^{-}$ can be written as
\cite{EPR2,DF}
\begin{eqnarray}
A(K_{L} \rightarrow \pi^{0} e^{+} e^{-}) & = & - \frac{G_{8} \alpha}{4 \pi}
(p + p^{\prime})^{\mu} {\bar u} [\phi_{V} \gamma_{\mu} + \phi_{A} \gamma_{\mu}
\gamma_{5}] v
\label{e50}
\end{eqnarray}
with
\begin{eqnarray}
\phi_{V} & = & \epsilon \phi_{S} - \frac{16 \pi^{2}}{3} i Im W_{1}
\label{e51}
\end{eqnarray}
and
\begin{eqnarray}
\phi_{A} & = & \frac{16 \pi^{2}}{3} i Im W_{5}.
\label{e52}
\end{eqnarray}
$W_{5}$ is defined by the effective Lagrangian counterterm
\begin{eqnarray}
{\cal L}_{w}^{\prime} & = & -\frac{i2 \pi \alpha}{3} G_{8} W_{5}
{\bar e} \gamma^{\mu} \gamma_{5} e Tr (\lambda_{6-i7}) (\partial U
- i e[A_{\mu},U])U^{\dag}.
\label{e53}
\end{eqnarray}
Taking the $\phi_{V}$ and $\phi_{A}$ as in \cite{DF,BLMM}
\begin{eqnarray}
\phi_{V} & = & e^{i \pi /4 } (0.57 \times 10^{-3}) - i 1.0 \times 10^{-3}
\label{e54}
\end{eqnarray}
and
\begin{eqnarray}
\phi_{A} & = & i 1.0 \times 10^{-3},
\label{e55}
\end{eqnarray}
we calculate  the branching ratio
$BR (K_{L} \rightarrow \pi^{0} e^{+} e^{-})$ for the counterterms couplings.
For $W_{+} = 1.22$ and  $W_{S} = 4.46$,
obtained in the case of the extended Nambu and  Jona - Lasinio model, we find
$BR (K_{L} \rightarrow \pi^{0} e^{+} e^{-})  =  1.15 \cdot 10^{-10}$
and for $W_{+} = 0.51$ and $W_{S} = 3.75$ the decay rate is
$BR (K_{L} \rightarrow \pi^{0} e^{+} e^{-})  = 7.77 \cdot 10^{-11}$.
{}From the large $N_{c}$ approach to the CP violating process we get
$BR (K_{L} \rightarrow \pi^{0} e^{+} e^{-})  = 4.79 \cdot 10^{-9}$.
Experimentally was determined   only the upper limit
$BR (K_{L} \rightarrow \pi^{0} e^{+} e^{-}) \leq  4.3 \cdot 10^{-9}$,
a result from \cite{DAH}, and
$BR (K_{L} \rightarrow \pi^{0} e^{+} e^{-}) \leq  4.3 \cdot 10^{-11}$
obtained by \cite{KEO}.
{}From our analysis it follows that indirect CP violation is rather
dependent on the $W_{S}$ counterterm coupling. It was noticed \cite{DF}
that it is very difficult to distinguish between direct and indirect
CP violation. These authors point out that the existence of direct
CP violating term in the branching
ratio $BR (K_{L} \rightarrow \pi^{0} e^{+} e^{-})$
can not be seen without observing branching ratio
$BR (K_{S} \rightarrow \pi^{0} e^{+} e^{-})$.
The CP conserving process proceeds via two photon exchange.
The contribution of vector and scalar mesons to $K_{L} \rightarrow \pi^{0}
\gamma \gamma$
is completely determined to ${\it O}(p^{6})$ \cite{EPR4,SF2}.
The authors of ref. \cite{DF} made a fit to the $\gamma \gamma$
spectrum in  $K_{L} \rightarrow \pi^{0}  \gamma \gamma$ in order to fix
the parameter describing the role of vector mesons at this order.
They obtain rather large value of the relevant parameter $a_{V}$,
comparing the results when the weak deformation
model was used like in \cite{EPR4,SF2}.  \\

{\bf 6 Summary}\\

We have investigated  counterterm couplings
required by CHPT in the decays $K^{+} \rightarrow \pi^{+} e^{+} e^{-} $,
$K_{S} \rightarrow \pi^{0} e^{+} e^{-} $ and
$K_{L} \rightarrow \pi^{0} e^{+} e^{-} $. The counterterms induced by the
strong,
weak and electromagnetic interactions  have been fixed by
resonance exchange. \\

- In our approach the weak deformation model, used for the
weak interactions, cannot reproduce the experimental result for
$K^{+} \rightarrow \pi^{+} e^{+} e^{-} $ decay rate. \\

- The branching ratio $BR(K^{+} \rightarrow \pi^{+} e^{+} e^{-}) $
and ${\hat c}$ defined in
$(K^{+} \rightarrow \pi^{+} \gamma \gamma)$ decay amplitude,
are in agreement with the
experimental results, when the factorization model combined with the set of
the parameters determined in the extended Nambu and Jona-Lasinio model is
applied.\\

- The counterterm couplings depend on a physically unobservable coupling
$H_{1}$.
At the leading order in large $N_{c}$  there is no dependence on the
$H_{1}$ coupling. \\

- The  $BR(K^{L} \rightarrow \pi^{0} e^{+} e^{-}) $  can be predicted
within these approaches. The decay rate is very parameter dependent.\\

{\bf Acknowledgements} I would like to thank E. de Rafael for a motivation
of this work and for many useful and stimulating discussions.
I also thank CNRS for the financial support during my visit to
University of Marseille.

\newpage
\begin{table}[h]
\begin{center}
\begin{tabular}{|c||c|c|c||c|c|c|}
\hline
$\nu$ & $m_{\rho}$ & $0.265 GeV$ & $1.165 GeV$ &
$m_{\rho}$ & $0.265 GeV $ & $1.165 GeV $ \\
\hline\hline
$W_{+}$ & $3.91$ & $3.20$ & $4.19$ & $2.69$ & $1.96$ & $4.19$  \\
\hline
$W_{S}$ & $3.49$ & $2.78$ & $3.77$ & $6.69$ & $5.98$ & $6.96$  \\
\hline
$c$ & $0$ & $0$ & $0$ & $-1.41$ & $-1.41$ & $-1.41$ \\
\hline
\end{tabular}
\end{center}
\caption{The counterterms $W_{+}$, $W_{S}$ calculated using  the extended
Nambu and Jona-Lasinio model for these values of the renormalization scale.
The first three columns contain the values
in the case of weak deformation model and in the last three columns
there are values obtained in the case of factorization model.}
\end{table}

\end{document}